# ENTREPRENEURIAL CAPABILITY AND ENGAGEMENT OF PERSONS WITH DISABILITIES TOWARD A FRAMEWORK FOR INCLUSIVE ENTREPRENEURSHIP

**XAVIER LAWRENCE D. MENDOZA, MBA**
Faculty Member - Department of Management Studies
Cavite State University - CCAT Campus
Rosario, Cavite, Philippines

## ABSTRACT

**The study was designed to determine the entrepreneurial capability and engagement of persons with disabilities towards a framework for inclusive entrepreneurship. The researcher used de- scriptive and correlational approaches through purposive random sampling. The sample came from the City of General Trias and the Municipality of Rosario, registered under their respec- tive Persons With Disabilities Affairs Offices (PDAO). The findings indicated that the respond- ents are from the working class, are primarily female, are mostly single, have college degrees, live in a medium-sized home, and earn the bare minimum. Furthermore, PWDs' perceived ca- pability level in entrepreneurship was somehow capable, and the majority of engagement level responses were somehow engaged. Considerably, age and civil status have significant relation- ships with most of the variables under study. Finally, the perceived challenges of PWDs' respondents noted the following: lack of financial capacity, access to credit and other financial institutions, absence of business information, absence of access to data, lack of compe- tent business skills, lack of family support, and lack of personal motivation. As a result, the au- thor proposed a framework that emphasizes interaction and cooperation between national and local government units in the formulation of policies promoting inclusive entrepreneurship for people with disabilities.**

## INTRODUCTION

Human productivity is very essential in nation building because it provides an avenue for society to grow economically and socially. Productivity is a widely held goal that society has attempted to strive for ages to achieve, and it is a determinant of wealth production (Lopez and Sy, 2014). Behaviorally speaking, personal productivity aids in the realization of an individual's self-worth or self -value. This transgresses through various types of working group because of the reward and remuneration factor that enables them to purchase or acquire (Flyndth, 2017). However, productivity and worth are sensitive issues when it comes to Persons with Disabilities (PWDs) because of certain constraints that limit their respective activities (Nicholas, 2012).

People with disabilities are people with long-term mental, sensory, and physical impairments who are constrained in their ability to perform activities and functions. PWDs were empowered in the Philippines through a variety of programs, activities, and legislative provisions ranging from social awareness, acceptance, and inclusivity (Guiterrez, 2013). However, PWDs are yet to be integrated into a sustainable economic platform that will enable them to provide for themselves, at least. PWDs are considered a marginalized sector that is continually submitted for continuous growth and developmental key identification. Entrepreneurship was recognized by the government as the primary strategy in poverty alleviation and mitigation as a powerful economic tool to foster a higher standard of living (Llanto, 2012).

Thus, the researchers attempted to understand and provide a focus on the entrepreneurial capability and engagement of PWDs to provide a pathway for possible sustainable livelihood for them, encourage policy makers and think-tanks to springboard entrepreneurial activities, and create inclusive programs and policies for the physically -challenged sector of our society.

## OBJECTIVES

The study aimed to answer the entrepreneurial engagement of persons with physical disability in City of General Trias and Municipality of Rosario, Cavite.

Specifically, the study aimed to:

1. determine the socio-economic profile of the respondents in terms of:
   a. age;
   b. sex;
   c. civil Status;
   d. educational attainment;
   e. household size; and
   f. combined household income.
2. determine the perceived capability level of the participants in entrepreneurial activity in terms of:
   a. business planning;
   b. marketing;
   c. financial planning;
   d. operation management; and
   e. leadership.
3. determine the perceived entrepreneurial engagement level in terms of:
   a. business planning;
   b. marketing;
   c. financial planning;
   d. operation management; and
   e. leadership.
4. determine the significant relationship between socio-economic profile and perceived capability level in entrepreneurial activities.
5. determine the significant relationship between socio-economic profile and perceived entrepreneurial engagement level.
6. identify the perceived challenges of the participants in engaging entrepreneurship.
7. propose an entrepreneurial framework for inclusive entrepreneurship of persons with disabilities in the province of Cavite.

## LITERATURE REVIEW

Persons with Disabilities (PWDs) belong to the marginalized sector of society. These people usually suffer from poor education and limited economic participation, resulting in a higher poverty rate compared to those without disabilities. Poverty issues for PWDs are a result of social exclusion and lack of opportunities in the mainstream population. It is true that everyone deserves to be treated fairly and respected, but in society, this situation is still a challenge for our brothers and sisters with disabilities. Limited access to education becomes a source of inequality for those who belong to marginalized groups, which consequently distresses their employability. Thus, this serves as a barrier for them to sustain their personal necessities and improve their social growth (World Health Organization, 2017), (Seyoum, 2017), and (Robertson and Bethea, 2018).

Awareness of different countries and private institutions arises regarding the exclusivity of PWDs. In the Philippines, one way of improving the welfare of PWDs is the passage of the Magna Carta for Persons with Disabilities. This promotes the anti-discrimination of PWDs by assuring their rights to employment, health, education, and auxiliary services. Part of the government's effort to promote inclusivity and equality among PWDs is the creation of an agency that can directly address the needs of disabled individuals. The Persons with Disability Affairs Office is an agency that implements services under the provisions of RA 7277. According to the studies of Mabaquiao (2018) and Macatangay (2018), different programs were being conducted in partnership with other government agencies and private institutions which mainly focused on health services and employment issues. This is also a prevalent scenario in the province of Cavite. PDAO in Cavite, together with local government units, strives to initiate sustainable programs for PWDs in Cavite, especially in the employment sector and other incentives.

Despite all the efforts of the government to uplift the social status of PWDs by providing opportunities related to employment, they still belong to the segment which requires relevant improvement. The study by Schelzig (2005) reveals that a small percentage of registered employable PWDs were employed. As a result, the majority of employed PWDs are employed in the agricultural sector (Mina, 2013), which significantly belongs to the country's marginalized group. Moreover, the occurrence of discrimination in the minds of Filipinos was not restrained. Proving this is the difficulty in searching for data on the impact of policies made for PWDs (Tabuga, 2013).

These ideas pushed the researcher to come up with this study. Since employment in the country is declining, this could give a pathway for a possible sustainable livelihood and enable strategies and policies to promote the welfare of PWDs. Thus, the researcher, based on the existing literature and studies, will give emphasis to understanding inclusive entrepreneurial activity, especially among PWDs, in General Trias City and the Municipality of Rosario, both in the province of Cavite. This research undertaking can springboard the opportunity for social inclusion of PWDs and can serve as a platform for sound, effective, and inclusive policies towards PWDs.

## METHODOLOGY

### Research Design

The researcher used descriptive and correlational approaches in the conduct of this study. Specifically, the researcher used descriptive design to describe the socio-economic profile of the respondents and likert-type measurements of capability level. Moreover, the study used a correlational-inferential design to determine the significant relationship between socio-economic profile and perceived level of entrepreneurship capabilities.

### Sampling Design

The researcher used non-parametric purposive sampling. The sample came from the City of General Trias and the Municipality of Rosario, registered under their respective Persons With Disabilities Affairs Offices (PDAO). The researcher used mental-capable respondents under PWDs. Initially, the researcher targeted 100 respondents from each research locale. However, due to the limitations and constraints brought by the pandemic, the researcher only retrieved and utilized 79 respondents from General Trias City and 48 respondents from the Municipality of Rosario.

### Research Instruments

The medium that was used in the study was a modified survey questionnaire form to gather the details of PWD participants, which was distributed in General Trias City and the Municipality of Rosario. The survey questionnaire was validated by experts composed of academicians and practitioners in business, entrepreneurship, and PWDs. Cronbach's alpha was calculated at 0.93 through SPSS, which verifies the reliability of the instrument-responses.

### Data Gathering Procedure

Primary data was used in this study through a modified survey questionnaire form. The questionnaire was validated and pretesting was conducted to assure the viability, validity, reliability and usability of the instruments.

Before the conduct of the study, the necessary consent and permissions were obtained from the office of the respective Local Government Unit to consider ethical issues. After obtaining consent and permission, the questionnaires were coordinated with the respective heads of the Persons with Disability Affairs Offices to be distributed among the registered PWDs under their jurisdiction. However, due to the limitations brought by the pandemic, about 79 respondents from General Trias City and 48 respondents from the Municipality of Rosario were being retrieved.

The secondary data came from research of related literature such as books, newspapers, and other related literature with regards to PWDs and entrepreneurship. Furthermore, academic online references were utilized. The author processed the data into tables and figures for analysis and interpretation.

### Statistical Treatment of Data

The researcher utilized the following statistical tests in this study:

1. descriptive designs were used in this study such as frequency, percentages, means and category counts.

   Formula: $P = \frac{f}{N} \times 100\%$

   Where:
   P = Percentage
   f = frequency
   n = sample size

2. Spearman Rank correlation was utilized through the use of Statistical Package for Social Science (SPSS) in determining the relationship of socio-economic profile and perceived capability level and engagement level of the respondents in entrepreneurial activities.

   $$p = 1 - \frac{6\sum d_t^2}{n(n^2-1)}$$

   Where:
   ***p*** = Spearman rank correlation
   **d** = the difference between the ranks

## FINDINGS

**Table 1. Profile of the Respondents According to Age**

| Age | Frequency | Percentage |
|---|---|---|
| 15 to 20 yrs old | 9 | 7.09 |
| 21 yrs old to 25 yrs old | 44 | 34.65 |
| 26 yrs old to 30 yrs old | 19 | 14.96 |
| 31 yrs old to 35 yrs old | 20 | 15.75 |
| 36 yrs old to 40 yrs old | 10 | 7.87 |
| 41 yrs old to 45 yrs old | 12 | 9.45 |
| 46 yrs old to 50 yrs old | 8 | 6.30 |
| 51 yrs old and above | 5 | 3.94 |
| **Total** | **127** | **100.00** |

Table 1 shows the age group of the PWD respondents. The table shows that most of the respondents belong to the age group of 21 years old to 25 years old, composing 44 of the total sample size with 34.65 percent, while the lowest belongs to the age group of 51 years old and above, with 3.94 percent accounting for 5 out of 127 participants.The observed age profile conforms to the data of the Philippine Statistics Authority, which indicates that more than 50 percent of the total household population of PWDs belong to the working age group.

**Table 2. Profile of the Respondents According to Sex**

| Sex | Frequency | Percentage |
|---|---|---|
| male | 48 | 37.80 |
| female | 79 | 62.20 |
| **Total** | **127** | **100.00** |

Table 2 shows the sex of the respondents. The majority of the PWDs respondents are female, with 62.20 percent constituting 79 of the total sample size. This implies that more female responded to the questionnaires compare to male. According to the most recent Philippine Statistics Authority data, males made up 50.9 percent of all PWD, while females made up 49.1 percent. According to these figures, there are 104 handicapped men for every 100 disabled women. Considerably, males with disabilities outnumbered females in the 0 to 64 age groups.

**Table 3. Profile of the Respondents According to Civil Status**

| Civil Status | Frequency | Percentage |
|---|---|---|
| Single | 76 | 59.84 |
| Married | 23 | 18.11 |
| Live-in | 12 | 9.45 |
| Widowed | 1 | 0.79 |
| Separated | 15 | 11.81 |
| **Total** | **127** | **100.00** |

Table 3 shows the respondents' civil status. Out of 127 responses, 76 are single, accounting for 59.87 percent of the total sample; 23 are married, accounting for 18.11 percent; 15 are separated, accounting for 11.81 percent; and 1 is widowed, accounting for 0.79 percent. This means that the majority of the respondents are single.

**Table 4 Profile of the Respondents According to Educational Attainment**

| Educational Attainment | Frequency | Percentage |
|---|---|---|
| Elementary undergraduate | 6 | 4.72 |
| Elementary graduate | 3 | 2.36 |
| High school undergraduate | 20 | 15.75 |
| High school graduate | 29 | 22.83 |
| College undergraduate | 27 | 21.26 |
| College graduate | 40 | 31.50 |
| Graduate Studies | 2 | 1.57 |
| **TOTAL** | **127** | **100.00** |

Table 4 shows the educational attainment of the respondents. This reveals that out of 127 respondents, 2 respondents (1.57 percent) finished graduate studies, 40 respondents (31.50 percent) possessed a college degree, 27 respondents (21.26 percent) were college undergraduates, 29 respondents (22.83 percent) finished high school, and 29 respondents (22.83 percent) were not able to complete basic education. These findings significantly show that the majority of PWD respondents were able to complete basic education and graduate from college as far as this study is concerned.

**Table 5 Profile of the Respondents According to Household Size**

| Household Size | Frequency | Percentage |
|---|---|---|
| 1 to 3 members | 49 | 38.58 |
| 4 to 6 members | 75 | 59.06 |
| 7 and above members | 3 | 2.36 |
| **Total** | **127** | **100.00** |

Table 5 shows the household size of the respondents. It reveals that the majority of the respondents belong to the group of four to six members, comprising 75 out of 127 with 59.06% of the total sample size, 49 of the respondents with 38.58 percent of the total sample belong to one to three members, and 3 respondents with 2.36 percent of the total sample belong to seven and above members. This suggests that the majority of the PWD respondents belong to an average household size.

**Table 6 Profile of the Respondents According to Combined Household Income**

| Combined Household Income | Frequency | Percentage |
|---|---|---|
| less than Php 10,000.00 | 86 | 67.72 |
| Php 10,001.00 to Php 20,000.00 | 22 | 17.32 |
| Php 20,001.00 to Php 30,000.00 | 13 | 10.24 |
| Php 30,001.00 to Php 40,000.00 | 6 | 4.72 |
| **Total** | **127** | **100.00** |

Table 6 shows the combined household income of the respondents. The majority of PWD respondents, 67.72 percent out of a total of 86, come from households with an income of less than Php 10,000.00. 17.32 percent of the total sample size (22 respondents) earn Php 10,001.00 to 20,000.00. Furthermore, 13 respondents, or 10.24 percent of the total sample size, earn Php 20,001.00 to Php 30,000.00, while 4.72 percent, or 6 respondents, earn Php 30,001.00 to 40,000.00.

## Perceived Capability and Engagement Level of the respondents in Entrepreneurial Activities

**Table 7 Summary of Perceived Capability Level**

| Category | Grand Mean | Descriptive Value |
|---|---|---|
| Business Planning | 3.83 | Somehow Capable |
| Marketing | 3.81 | Somehow Capable |
| Financial Planning | 3.80 | Somehow Capable |
| Operational Management | 3.90 | Somehow Capable |
| Leadership | 3.91 | Somehow Capable |

Table 7 showcases the summary of the perceived capability level of the PWDs respondents. This explicitly shows that perceived capability in leadership is the most considerable activity among others with the grand mean of 3.91, while the least is capability in financial planning with grand mean of 3.80. Generally, most of the respondents appeared to be somehow capable of entrepreneurial activities. Disability, according to Norafandi, et. al. (2017)'s study on The Prospects of People with Disabilities Participating in Entrepreneur- ship, is a phenomena that occurs in the public view. Individuals with disabilities are frequently pushed to grasp ableism, to resemble those who are capable, in order to overcome their circumstances.

**Table 8 Summary of Potential Entrepreneurial Engagement Level**

| Category | Grand Mean | Descriptive Value |
|---|---|---|
| Business Planning | 3.85 | Somehow Engaged |
| Marketing | 3.81 | Somehow Engaged |
| Financial Planning | 3.83 | Somehow Engaged |
| Operational Management | 3.80 | Somehow Engaged |
| Leadership | 3.87 | Somehow Engaged |

Table 8 shows the summary of the entrepreneurial engagement level of PWD respondents. Among the entrepreneurial activities, leadership presents the highest nominal value with the grand mean of 3.87. This means that among engaging entrepreneurial activities, leadership is the most considerable aspect for PWD respondents, and apparently the least considerable is operational management with the grand mean of 3.80. Furthermore, this suggests that PWDs are somehow engaged in entrepreneurial activities. Considerably, these results are supported by the study of Norafandi, et.al. (2017) on social business which explains that engaging in business expanded PWDs' self-force and independence. Furthermore, investment in businesses reinforced their confidence, autonomy, and lessened the inclination that inabilities are signs of their lack.

## Perceived Challenges of the respondents in Engaging Entrepreneurship

**Table 9 Perceived Challenges in Doing Business**

| Challenges | Frequency | Percentage |
|---|---|---|
| Lack of financial capacity | 58 | 38.41 |
| Access to credit and other financial institution | 25 | 16.56 |
| Absence of business information | 25 | 16.56 |
| Absence of access to data | 9 | 5.96 |
| Not competent business skills | 13 | 8.61 |
| Lack of family support | 4 | 2.65 |
| Lack of personal motivation | 5 | 3.31 |
| Others | 12 | 7.95 |
| **Total** | **151** | **100.00** |

Table 9 reveals the perceived challenges of PWD respondents in doing business. This shows that lack of financial capacity, with 38.41 percent, composing 58 of the total sample size, access to credit and other financial institutions, and absence of business information, both with 16.56 percent composing of 25 responses out of the total sample, are the primary perceived challenges of PWDs in venturing into business. Anyone attempting to start a new firm faces a significant risk of encountering obstacles. Such limits become increasingly defined and complicated for individuals with disabilities. The results affirm the study by Cooney (2008) in which starting up a business is normally difficult for both disabled and non-disabled people. Moreover, Dhar and Farzana (2017) explain that social perceptions and financial aspects are the factors that establish the boundaries for those people with incapacities toward entrepreneurship.

**Relationship Between Socio - Economic Profile and Perceived Engagement and Capability Level of the Participants in Entrepreneurial Activity.**

**Table 10. Summary of Correlation**

| Category | General Significance ( p-value, crit < 0.05 ) | Null Hypothesis |
|---|---|---|
| Age to perceived capability | Significant | Reject |
| Age to perceived engagement | Significant | Reject |
| Sex to perceived capability | Insignificant | Accept |
| Sex to perceived engagement | Insignificant | Accept |
| Civil Status to perceived capability | Significant | Reject |
| Civil Status to perceived engagement | Significant | Reject |
| Educational Attainment to perceived capability | Insignificant | Accept |
| Educational Attainment to perceived engagement | Insignificant | Accept |
| Household Size to perceived capability | Insignificant | Accept |
| Household Size to perceived engagement | Significant | Reject |
| Household Income to perceived capability | Insignificant | Accept |
| Household Income to perceived engagement | Insignificant | Accept |

NB: general significance are calibrated by the greater composite value of the totality of coefficients and p-value of variables under study

Table 10 shows the summary of the correlation of the variables under study. Generally, the study reveals that sex, educational attainment, household size, and combined household income are insignificant, respectively. Thus, the null hypothesis is accepted. On the other hand, age and civil status have a significant relationship with perceived capability. Therefore, the null hypothesis is rejected. PWD individuals become more capable in business as they age, and their civil status has a major impact on this. Accepting the null hypothesis, sex, educational attainment, and household income are all insignificant predictors of entrepreneurial engagement.While age, civil status, and household size all have a significant impact on perceived entrepreneurial engagement, As a result, the null hypothesis is rejected. This suggests that PWD respondents are more likely to engage in business as they get older, and that, like perceived capability, civil status has a significant impact on entrepreneurial engagement, whereas the larger the respondents' household size, the less likely they are to engage in business activities.

## CONCLUSION

Based from the findings, the following conclusions can be stated:

1. The respondents are from the working class, dominated by women, mostly single, have college degrees, belong to an average-sized house-hold, and belong to a minimum-earning household.
2. The perceived capability level of PWDs in business planning, marketing, financial planning, operation management, and leadership are somehow capable, which means PWD respondents are capable of entrepreneurship. Considerably, the potential engagement level of PWDs in entrepreneurship in terms of business planning, marketing, financial planning, operation management, and leadership is somehow engaged, which means that PWD respondents are engaged in entrepreneurship.
3. Generally, age and civil status have significant relationships with most of the variables under study. Considerably, household size and entrepreneurial engagement have a significant relationship with perceived capability, which has an insignificant relationship. Furthermore, the majority of the variables under study have insignificant relationships with sex, educational attainment, and combined household income.
4. The perceived challenges of PWDs respondents noted the following: lack of financial capacity, access to credit and other financial institutions, absence of business information, absence of access to data, lack of competent business skills, lack of family support, lack of personal motivation, and others (competition in the market and a diverse workforce).

## RECOMMENDATION

Based on findings and conclusions of the study, the following recommendation are given:

1. To ensure certainty of access to financial assistance for starting a business, public policy must ensure that potential PWDs can obtain financial assistance for business creation.
2. Local government units should assist PWDs in obtaining equal rights and opportunities in obtaining financial resources from financial institutions for their business development, while banking institutions should also allow financing programs for PWDs that could be used in funding their viable businesses.

3. Under the supervision of the Local Government unit, the PDAO should give and facilitate information in an accessible format while also ensuring that these financial schemes do not discriminate against people with disabilities.
4. To raise awareness of the potential entrepreneurship for PWDs, Social Welfare Development should raise awareness of self-employment and small business ownership as a potential advantage for self-supporting activity and becoming self-reliant.
5. In order to bridge the experience gap among PWDs, the local government unit, in collaboration with the Department of Trade and Industry, may promote the acquisition of entrepreneurial skills.
6. To encourage participation of private business sectors, the Local Government Unit might speak with these groups and persuade them to allocate a portion of their business's production and distribution process.
7. The proposed framework for inclusive entrepreneurship may be considered in this study.

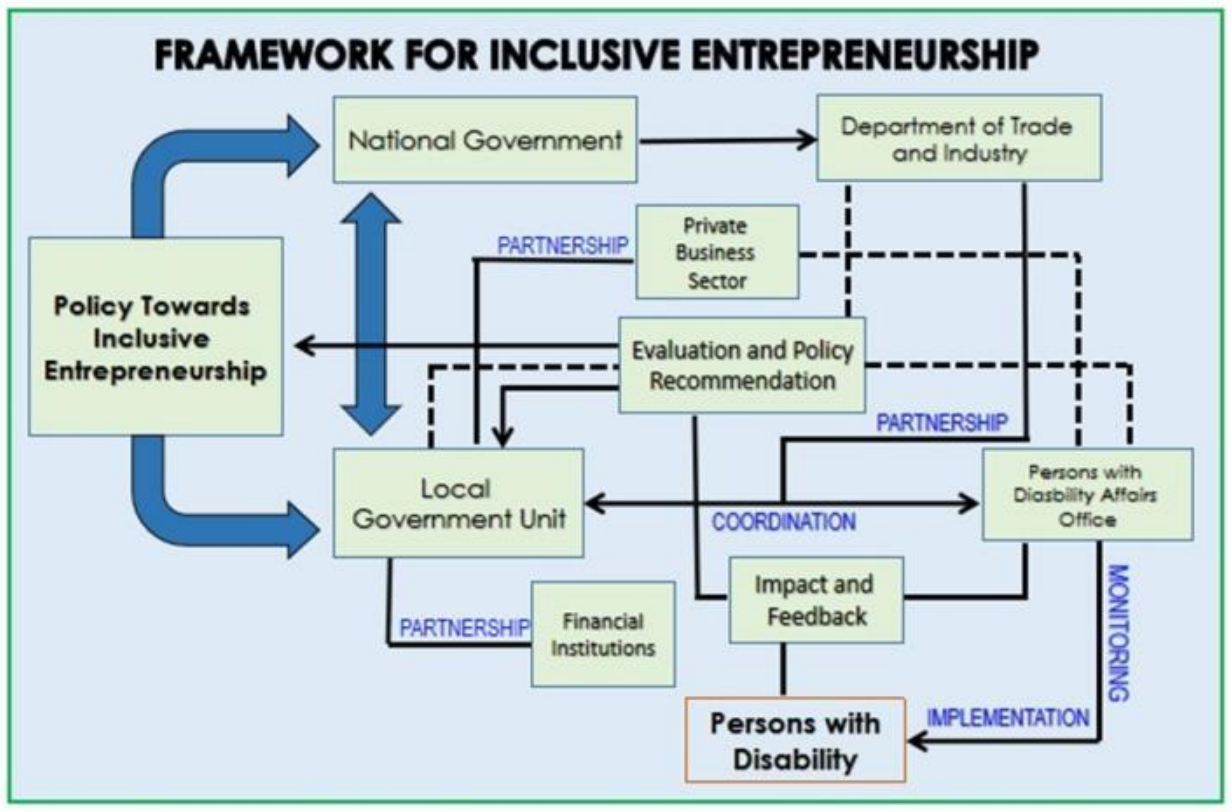

**Figure 1. Proposed Framework of Inclusive Entrepreneurship for PWDs**

This framework reveals that the National Government and Local Government Units should initiate the formulation of policy towards inclusive entrepreneurship for PWDs. Under the National Government, the Department of Trade and Industry may enter into partnership with the Local Government Unit and Persons with Disability Affairs Office in promoting and providing entrepreneurial skills development as well as business information to potential PWD entrepreneurs. Moreover, local government units may tap financial institutions within their jurisdictions to provide financing programs for developing businesses. At the same time, the private sector may be encouraged to include capable PWDs in business activities through the supervision of PDAO. The implementation and monitoring of such a policy will be under the PDAO since their office has direct engagement with persons with disabilities. Thus, the impact on PWDs of the adoption of entrepreneurial policies as well as feedback will be assessed and processed by PDAO, Local Government Unit, and Department of Trade and Industry for gap identification, which will be the basis for the continuous development of persons with disabilities in entrepreneurship.

## ACKNOWLEDGEMENT

The researcher is extremely grateful to God Almighty since this study would not have been feasible without His favors and blessings. Ms. Janica Tabbu, Mr. Jerico Tadeo, Ms. Nerisa Abug, and Mr. Richard Perez have the author's undying thanks for their assistance and support.